\ifx\fiverm\undefined
        \newfont\fiverm{cmr5}
\fi
\input prepictex
\input pictex
\input postpictex
\catcode`\@=11
\@input{picmore.tex}
\documentstyle[11pt]{article}

\def\labelmark{}

\def\void{}
\newenvironment{formula}[1]{\def\labelname{#1}
\ifx\void\labelname\def\junk{\begin{displaymath}}
\else\def\junk{\begin{equation}\label{\labelname}}\fi\junk}%
{\ifx\void\labelname\def\junk{\end{displaymath}}
\else\def\junk{\end{equation}}\fi\junk\labelmark\def\labelname{}}

{\ifx\void\labelname\def\junk{\end{array}\end{displaymath}}
\else\def\junk{\end{array}\right.\end{equation}}
\fi\junk\labelmark\def\labelname{}\def\junk{}
}

\newcommand{\beq}{\begin{formula}}
\newcommand{\eeq}{\end{formula}}
\newcommand{\beqa}{\begin{eqnarray}}
\newcommand{\eeqa}{\end{eqnarray}}
\newcommand{\eq}[1]{(\ref{#1})}
\newcommand{\nn}{\nonumber}

\newcommand{\ra}{\rightarrow}
\newcommand{\der}{\partial}

\newcommand{\NP}[1]{ {\it Nucl.~Phys.} {\bf #1}}
 
\newcommand{\PL}[1]{ {\it Phys.~Lett.} {\bf #1}}

\newcommand{\PR}[1]{ {\it Phys.~Rev.} {\bf #1}}

\newcommand{\AP}[1]{ {\it Ann.~Phys.} {\bf #1}}

\newcommand{\ACT}[1]{ {\it Acta~Phys.~Polon.} {\bf #1}}
\newcommand{\EPJ}[1]{ {\it Europ.~Phys.~J.} {\bf #1}}
\newcommand{\JMP}[1]{ {\it J. Math.~Phys.} {\bf #1}}

\begin{document}
\begin{titlepage}
\setcounter{page}{0}
\renewcommand{\thefootnote}{\fnsymbol{footnote}}

\begin{flushright}
\mbox{\phantom{draft v3}} \\
\mbox{\phantom{HD-THEP-98}} \\
\end{flushright}

\vspace{5 mm}
\begin{center}
{\Large\bf World-line Green functions with\\
momentum and source conservations} 

\vspace{10 mm}

{\bf Haru-Tada Sato$^{}$
\footnote{E-mail: sato@thphys.uni-heidelberg.de }}
\vspace{5mm}

\vspace{10mm}

{\it $^{}$ Institut f{\"u}r Theoretische Physik\\
Universit{\"a}t Heidelberg\\
Philosophenweg 16, D-69120 Heidelberg, Germany}
\end{center}

\vspace{10mm}
\begin{abstract}
Based on the generating functional method with an external 
source function, a useful constraint on the source function 
is proposed for analyzing the one- and two-loop world-line 
Green functions. The constraint plays the same role as the 
momentum conservation law of a certain nontrivial form, 
and transforms ambiguous Green functions into the 
uniquely defined Green functions. We also argue 
reparametrizations of the Green functions defined on 
differently parameterized world-line diagrams. 
\end{abstract}

\vfill

\begin{flushleft}
\end{flushleft}
\end{titlepage}
%\newpage
%%%%%%%%%%%%%%%%%%%%%%%%%%%%%%%%%%%%%%%%%%%%%%%%%%%%%%%%%%%%%%%%%%%%%
\setcounter{footnote}{0}
\renewcommand{\thefootnote}{\arabic{footnote}}
\renewcommand{\theequation}{\thesection.\arabic{equation}}
%\renewcommand{\theequation}{\arabic{equation}}
%--------------------------------------------------------------------
\section*{I. Introduction}\label{sec1}
\setcounter{section}{1}
\setcounter{equation}{0}
\indent
%--------------------------------------------------------------------

String theory organizes the scattering amplitudes in a very 
compact form (in the infinite string tension limit), and this fact 
makes the investigation of field theory scattering amplitudes very 
nontrivial and potentially useful~\cite{BK}-\cite{world}. 
In this spirit, multi-loop scattering amplitudes have also been 
studied both from the string theory viewpoint~\cite{multi}-\cite{RS2} 
and the field theory based on the first quantization formalism 
(world-line formalism)~\cite{SSqed}-\cite{MSS2}. 
The general structure of field theory 
amplitudes (with $N$ external momenta $p_1,p_2,\cdots,p_N$) at the 
one-loop order is described as~\cite{BK,str} 
\beq{mas}
\Gamma_N=\int_0^\infty {dT\over T}({1\over4\pi T})^{D\over2}
(\prod_{n=1}^N\int_0^Td\tau_n) K(\tau_1,\tau_2,\cdots,\tau_n;T)
\exp[{1\over2}\sum_{j,k=1}^Np_j\cdot p_k G_B(\tau_j,\tau_k)]\ ,
\eeq
where $K$ is a certain function which depends on the detail of 
theory in question (it can be determined systematically). 
The exponent including the function $G_B$ is sometimes called 
the generating kinematical factor, and is a theory independent 
object. $G_B$ is the (world-line) Green function between two 
points on a loop of length $T$. One can similarly write down the 
generalized formulae for certain multi-loop cases~\cite{RS2,MSS} 
with using multi-loop Green functions~\cite{RS1,SSphi}. In this 
sense, determinations of multi-loop Green functions are important 
factors in the world-line formalism.   

In this paper, we focus on an ambiguity problem of the 
multi-loop Green functions. There is an ambiguity involved 
in the world-line Green functions, raised from the momentum 
conservation law. All of such Green functions should be 
reduced to the uniquely defined ones under the constraint 
of a vanishing identity, without changing the value of a 
kinematical factor. The problem is trivial in the one-loop 
case, and summarized as follows. In the original definition
\beq{*1}
{1\over2}\der^2 G_B(\tau)=\delta(\tau)-{1\over T}
\eeq
with imposing rotational invariance and periodicity, $G_B$ 
is uniquely determined as the rotational symmetric form 
\beq{gb}
G_B(\tau_1,\tau_2) = G_B(\tau_1-\tau_2)
=|\tau_1-\tau_2|-{(\tau_1-\tau_2)^2\over T}\ .
\eeq
However, we do not necessarily use this functional form as concerns 
the kinematical factor itself, actually which does not change 
if we add a polynomial in $\tau_j$ to the $G_B(\tau_j,\tau_k)$ in 
\eq{mas} because of the conservation law $\sum_k p_k=0$. 
This ambiguity does not cause any problem in the one-loop case, 
where the defining equation of $G_B$ is simple and its rotational 
invariance should be clear. Nevertheless, in multi-loop cases, 
it is certainly useful to identify the ambiguity which can be 
canceled by a certain condition such as conservation law, because 
the definitions and calculations of multi-loop Green functions are 
in general complicated --- in addition that the rotational 
invariance is unclear in certain cases. 

As the simplest nontrivial example, we discuss the two-loop 
Green functions~\cite{SSphi,HTS}. We refer to the Green functions 
containing the ambiguity as the {\it wide sense} Green functions. 
As suggested above, the value of a kinematical factor should be 
invariant for any set of wide sense Green functions. Various wide 
sense Green functions can be obtained depending on how to define 
and evaluate, and we shall verify that all of them can be identified 
with each other in the sense of keeping the kinematical factor 
invariant by way of examples. To this end, we obviously need a 
constraint such as the total momentum conservation law. 
However, in the generic multi-loop cases, a useful form of the 
conservation law is in practice not a simple summation 
(along the single loop as mentioned in the one-loop case), 
because of the presence of additional internal lines. 
We generalize the conservation law into a more suitable form 
to our purpose. In addition to the momentum conservation law, 
we also present a continuous analogue of the conservation law; 
that is a constraint on the integrals of external source 
functions along the two-loop world-line vacuum diagram. 
This continuous version is very simple and useful to 
apply practical computations, and is non-trivial since 
such a constraint does not exist in the source term of 
the usual formulation of field theory. 

In Section~II, for notational conveniences, we briefly 
review the two-loop kinematical factor and the Green functions 
in $\phi^3$ theory. In Section~III, we present a 
useful two-loop momentum conservation formula, and demonstrate 
how to apply the formula to the identification of different 
wide sense Green functions. In Section~IV, employing 
the generating functional method with external source 
functions~\cite{MSS2}, we consider another derivation of the 
Green functions. In this case, we show that there also exists a 
similar constraint formula on the source functions, and verify 
that it plays the same role as the momentum conservation method 
of Section~III. In Section~V, we further confirm 
the validity and usefulness of the source constraint 
in more specific cases (one-loop QED). Section~VI 
is a short note on the previous work~\cite{HTS}, 
concerning new reparametrization 
transformations of the two-loop Green functions. 
Section~VII contains conclusions. 

%--------------------------------------------------------------------
\section*{II. Notations}\label{sec2}
\setcounter{section}{2}
\setcounter{equation}{0}
\indent
%--------------------------------------------------------------------
{}~For the purpose of setting our notations, we briefly review 
the world-line Green functions and the master amplitude formulae 
corresponding to Eq.\eq{mas} in the two-loop $\phi^3$ 
theory~\cite{RS2,SSphi,MSS}. The (two-loop) master formula is a 
fundamental quantity which contains all necessary Feynman diagrams 
belonging to a certain class of diagrams. The classes are labeled 
by two or three integers $(N',N_3)$ or $(N_1,N_2,N_3)$, and amplitudes 
are certain combinations of these classes~\cite{MSS}. We call 
the first labeling the loop type, and the latter the symmetric 
type. The general form of the master formula is as follow:
\beq{master}
\Gamma_M^{2-loop}= {1\over12}(-g)^{N+2}
\int dM\, (4\pi)^{-D}\Delta^{D\over2} \exp[\,E_G\,]\ .
\eeq
{}~For the loop type parameterization ($N=N'+N_3$), 
the integration measure $dM$ is 
\beq{*2}
dM=  {dT\over T} dT_3 d\tau_\alpha d\tau_\beta 
    \prod_{n=1}^{N'} d\tau_n \prod_{l=1}^{N_3}d\tau^{(3)}_l\ ,
\eeq
and $\Delta$ is the determinant factor 
\beq{*3}
\Delta=\Bigl(\,TT_3+TG_B(\tau_\alpha,\tau_\beta)\,\Bigr)^{-1}\ .
\eeq
The exponential part $E_G$, the generating kinematical factor, 
takes the following bilinear form in $N$ external momenta 
($p_j,p^{(3)}_k; j=1,\cdots,N';k=1,\cdots,N_3$): 
\beq{gkf1}
E_G={1\over2}\sum_{j,k=1}^{N'} p_j p_k G^{(1)}_{00}(\tau_j,\tau_k)
     +{1\over2}\sum_{j,k=1}^{N_3} p^{(3)}_j p^{(3)}_k 
                    G^{(1)}_{33}(\tau^{(3)}_j,\tau^{(3)}_k)
     +\sum_{j=1}^{N'}\sum_{k=1}^{N_3} p_j p^{(3)}_k 
                    G^{(1)}_{03}(\tau_j, \tau^{(3)}_k)  \ ,  
\eeq
where the bilinear momenta should be understood as Lorentz contracted 
forms (hereafter as well). 
The explicit forms of these Green functions are~\cite{HTS} 
\beq{g00}
G^{(1)}_{00}(\tau,\tau')=G_B(\tau,\tau')-{1\over4}{ 
  \left( G_B(\tau,\tau_\alpha) - G_B(\tau,\tau_\beta)
   - G_B(\tau',\tau_\alpha) + G_B(\tau',\tau_\beta)\right)^2
      \over T_3+G_B(\tau_\alpha,\tau_\beta)}          \label{g00}
\eeq
\beq{g11}
G^{(1)}_{33}(z_1,z_2)=G^{(1)}_{33}(z_1-z_2)=
=\vert z_1-z_2\vert   -{(z_1-z_2)^2\over 
T_3+G_B(\tau_\alpha,\tau_\beta)},  \label{g11}
\eeq
\beq{g01}
G^{(1)}_{03}(\tau,z)= \left\{\begin{array}{ll} 
G^{(1)}_{00}(\tau,\tau_\alpha)+\frac{1}
{T_3+G_B(\tau_\alpha,\tau_\beta)}
\left(T_3z- z^2+ z[G_B(\tau,\tau_\beta)-G_B(\tau,\tau_\alpha)]\right)
&\quad\mbox{for}\quad \tau_\beta < \tau_\alpha      \\
G^{(1)}_{00}(\tau,\tau_\beta)+\frac{1}{T_3 
+ G_B(\tau_\alpha,\tau_\beta)}
\left(T_3z- z^2 + z[G_B(\tau,\tau_\alpha) 
-G_B(\tau,\tau_\beta)]\right)
&\quad\mbox{for}\quad \tau_\alpha < \tau_\beta
\end{array}\right.                                 \label{g01}
\eeq
The $\tau$ parameters $\{\tau_\alpha,\tau_\beta,\tau_n|n=1,
\cdots,N'\}$ run from zero to $T$, which stands for the length 
of a loop (fundamental loop), and $\tau^{(3)}_n$, $n=1,\cdots,N_3$ 
run from zero to $T_3$, the length of the internal line (the rest 
part of the vacuum diagram). $T$ and $T_3$ are to be integrated 
from zero to infinity. In \cite{RS2}, we pointed out that one 
may fix and eliminate one of the parameters $\{\tau_\alpha,
\tau_\beta,\tau_n|n=1,\cdots,N'\}$ because of the rotational 
symmetry of the fundamental loop. This means that we can set one 
of these parameters to be zero which corresponds to the origin of 
world-line coordinate along the fundamental loop. Obviously, 
$G^{(1)}_{00}$ is invariant under this rotation, and does not 
receive any serious change, however $G^{(1)}_{03}$ does not even 
possess any translational symmetry such as seen in $G^{(1)}_{33}$. 
Hence the explicit form of $G^{(1)}_{03}$ depends on which 
parameter will be set zero. For example, if we choose $\tau_\beta$ 
as such origin, $G^{(1)}_{03}$ should follow the form for 
$\tau_\beta < \tau_\alpha$. Similarly, if $\tau_\alpha$, then 
take for $\tau_\alpha < \tau_\beta$. There is also a different Green 
function~\cite{SSphi} from Eq.\eq{g00}. However, both coincide 
under the same momentum conservation constraint (for $N_3=0$) 
as the one-loop type.

Using the transformation obtained in Ref.~\cite{HTS}, we can 
transform the above quantities to the other version (symmetric 
parameterization). It is done by dividing the fundamental loop 
into two pieces $T=T_1+T_2$ with $N'=N_1+N_2$ and 
$\{\tau_n\}\ra\{\tau^{(1)}_n,\tau^{(2)}_n\}$. 
In this case, we have~\cite{SSphi} 
\beqa
dM &=&  dT_1dT_2dT_3
    \prod_{i=1}^3 \prod_{n=1}^{N_i}d\tau_n^{(i)}\ , \\
\Delta&=& (T_1T_2+T_2T_3+T_3T_1)^{-1}\ ,
\eeqa
and
\beq{gkf2} 
E_G={1\over2}\sum_{a=1}^3 \sum_{j,k}^{N_a} p^{(a)}_j p^{(a)}_k 
           G^{\rm sym}_{aa}(\tau^{(a)}_j,\tau^{(a)}_k)
   +\sum_{a=1}^3\sum_j^{N_a}\sum_k^{N_{a+1}} p^{(a)}_j p^{(a+1)}_k 
           G^{\rm sym}_{aa+1}(\tau^{(a)}_j,\tau^{(a+1)}_k) , 
\eeq
where we set $\tau^{(4)}=\tau^{(1)}$ and $N_4=N_1$ etc. in accord 
with the cyclic expression. The Green functions are~\cite{SSphi}
\beq{gsymaa}
G^{\rm sym}_{aa}(\tau,\tau')=G^{\rm sym}_{aa}(\tau-\tau')
=\vert \tau-\tau'\vert -{T_{a+1}+T_{a+2}\over 
           T_1T_2+T_2T_3+T_3T_1}(\tau-\tau')^2,    \label{gsymaa}
\eeq
\beq{gsymaa1}
G^{\rm sym}_{aa+1}(\tau,\tau')=\tau+\tau'- 
{\tau^2T_{a+1}+{\tau'}^2T_{a}+(\tau+\tau')^2T_{a+2} \over
                      T_1T_2+T_2T_3+T_3T_1}.     \label{gsymaa1}
\eeq
All the formulae in this section are reproduced from string 
theory~\cite{RS1,RS2}, and in this sense, we refer to these 
Green functions \eq{g00}-\eq{g01}, \eq{gsymaa}, and \eq{gsymaa1} 
as the standard forms. 

%--------------------------------------------------------------------
\section*{III. Momentum conservation constraint}\label{sec3}
\setcounter{section}{3}
\setcounter{equation}{0}
\indent
%--------------------------------------------------------------------

In this section, 
we encounter the (wide sense) Green functions of different 
forms, depending on calculation methods (in the symmetric 
parameterization). However, the value of $E_G$ should be shown to be 
invariant under the constraint of total momentum conservation. 
In the one-loop case, as mentioned in the introduction, 
the constraint is expressed by the identity
\beq{mom1}
\sum_{j=1}^N\sum_{k=1}^Np_j\cdot p_k\tau^m_k=0\ .
\eeq
The single summation over all momenta $p_j$ is nothing but the 
summation over the fundamental loop. However, the same structure 
can not be seen in \eq{gkf1} or \eq{gkf2} for the $N_3\not=0$ case. 
Hence, we shall derive a suitable two-loop generalization of 
this identity, and explain how it works. To illustrate the idea 
clearly, we need a couple of examples of different Green 
functions in the first place. 

As explained in Ref.~\cite{MSS}, Eq.\eq{master} is obtained from 
the path integral 
\beqa
\Gamma^{2-loop}_M &=&
{(-g)^{N+2}\over2\cdot3!} 
\int d^Dx_1d^Dx_2 \prod_{a=1}^3\int_0^\infty dT_a e^{-m^2T_a}\nn\\
&\times&\int_{\scriptstyle y_a(0)=x_2 
        \atop\scriptstyle y_a(T_a)=x_1}{\cal D}y_a(\tau)
\exp\Bigl[-\int_0^{T_a}{1\over4}{\dot y}_a^2 d\tau^{(a)}\Bigr]
\prod_{n=1}^{N_a} \int_0^{T_a}d\tau_n^{(a)}
e^{ip_n^{(a)}y(\tau_n^{(a)})}  \label{Gammaplanenew}
\eeqa
by using the mode expansion
\beq{*4}
y_a(\tau)=x_1+{\tau\over T_a}(x_2-x_1)+\sum_{m=1}^\infty
y_m {\rm sin}\Bigl({m\pi\tau\over T_a}\Bigr) \ . 
\eeq
A straightforward computation in this case shows that the $E_G$ 
part is composed of the following Green functions instead of 
$G^{\rm{sym}}_{ab}$
\beqa
G^M_{aa}(\tau,\tau')&=&|\tau-\tau'|-(\tau+\tau')
+2{\tau\tau'\over T_a}(1-\Delta {T_1T_2T_3\over T_a})\ ,\\
G^M_{aa+1}(\tau,\tau')&=&-2\Delta T_1T_2T_3
{\tau\tau'\over T_aT_{a+1}} \ .
\eeqa
Note that the $x_1$ integration generates the total momentum 
conservation factor 
\beq{delta}
(2\pi)^D\delta(\sum_a\sum_n^{N_a} p_n^{(a)}). 
\eeq

A second example is from Ref.~\cite{MSS2}. 
The world-line Green function should also be derived as a 
two-point function in the sense of ordinary field theory: 
\beq{*5}
{\cal G}_{\mu\nu}(\tau^{(a)}_1,\tau^{(b)}_2) =
<x_\mu(\tau^{(a)}_1)x_\nu(\tau^{(b)}_2)> = 
{\delta\over\delta J^\mu_a(\tau^{(a)}_1)}
{\delta\over\delta J^\nu_c(\tau^{(b)}_2)}
\ln Z[J]\Bigr|_{J=0} \ ,
\eeq
where the generating functional is given by 
\beq{*6}
Z[J]\equiv\int d^Dy_1 d^Dy_2 \Bigl(\prod_{a=1}^3 \,
\int_{\scriptstyle x_a(T_a)=y_2 
        \atop\scriptstyle x_a(0)=y_1}{\cal D}x_a\Bigr)
\exp\Bigl[-{1\over4}\sum_a\int_0^{T_a}{\dot x}^2_a(\tau)\,d\tau +
\sum_a\int_0^{T_a} J_a^\mu(\tau)x_a^\mu(\tau) d\tau \,\Bigr]\ .
\eeq
{}~For later convenience, we here write the intermediate 
expression (putting $w={y_1+y_2\over2}, z=y_2-y_1$) 
\beqa
Z[J]&=&\Bigl(\Pi_{a=1}^3(4\pi T_a)^{-D/2}\Bigr)
\exp[-{1\over2}\sum_{a=1}^3\int_0^{T_a}\int_0^{T_a}
J^a_\mu(\tau){\tilde G}^{(a)}_{\mu\nu}(\tau,\tau')
J^a_\nu(\tau')d\tau d\tau'] \nn\\
&\times&\int dz dw\prod_{a=1}^3\exp[w\int_0^{T_a}J^a(\tau)d\tau
-{1\over4}z^\mu A^a_{\mu\nu}z^\nu +z^\nu\int_0^{T_a}
J^a_\mu R^a_{\mu\nu}\,] \label{inter}
\eeqa
as well as the final expression 
\beqa
Z[J] &=& i\delta^D(\sum_{a=1}^3\int_0^{T_a} J_a^\mu(\tau)d\tau )
(4\pi)^{D\over2}
(\prod_{a=1}^3 (4\pi T_a)^{-D\over2})
\mbox{Det}_L^{-{1\over2}}(\sum_a A^a) \nn\\ 
&&\times \exp\Bigl[\,-{1\over2}\sum_a \int_0^{T_a}\int_0^{T_a}
J^a_\mu(\tau) {\tilde G}^{(a)}_{\mu\nu}(\tau,\tau') J^a_\nu(\tau') 
d\tau d\tau'\,\Bigr] \nn\\
&&\times \exp\Bigl[\, (\sum_a A^a)^{-1}_{\rho\sigma}\,
(\,\sum_a \int_0^{T_a} R^a_{\rho\mu} J^a_\mu(\tau) d\tau \,)
(\,\sum_c \int_0^{T_c} R^c_{\sigma\nu} J^c_\nu(\tau) d\tau\,)
\,\Bigr] \ , \label{inter2}
\eeqa
where we take
\beq{AR}
A^a_{\mu\nu}=\delta_{\mu\nu}T^{-1}_a\ ,\qquad
R^a_{\mu\nu}=({\tau^{(a)}\over T_a}-{1\over2})\delta_{\mu\nu}\ ,
\eeq
and
\beq{*7}
{\tilde G}^{(a)}_{\mu\nu}(\tau_1,\tau_2)=\delta_{\mu\nu}
\Bigl(|\tau_1-\tau_2|-(\tau_1+\tau_2)
+2{\tau_1\tau_2\over T_a}\Bigr)\ .
\eeq
A main difference from the first example is the existence of 
non-constant source terms, where further $\tau$ integrations are 
formally impossible. This is the reason of having a different 
form of Green function. Further decoupling the metric factor 
($g_{\mu\nu}=-\delta_{\mu\nu}$)
\beq{*8}
{\cal G}_{\mu\nu}(\tau^{(a)},\tau^{(b)})
=-g_{\mu\nu}G^J_{ab}(\tau^{(a)},\tau^{(b)})\ ,
\eeq
we actually derive the second different form
\beq{GJ}
G^J_{ab}(\tau^{(a)}_1,\tau^{(b)}_2) 
=\delta_{ab}
\Bigl(|\tau^{(a)}_1-\tau^{(b)}_2|-(\tau^{(a)}_1+\tau^{(b)}_2)
+2{\tau^{(a)}_1\tau^{(b)}_2\over T_a}\Bigr)
-{1\over2}(2\tau^{(a)}_1-T_a)(2\tau^{(b)}_2-T_b)
{T_1T_2T_3\over T_a T_b}\Delta \ .
\eeq

Now, let us derive the two-loop version of the 
constraint \eq{mom1}. It can be derived by combining the 
trivial identities
\beq{*9}
(\sum_{a=1}^3\sum_{j=1}^{N_a}p^{(a)}_j)
\sum_{k=1}^{N_b}p^{(b)}_k(\tau^{(b)}_k)^m=0\ ; 
\qquad b=1,2,3
\eeq
with multiplying weight coefficients $C^{(b)}_m$. The result is 
arranged in the form suitable to $E_G$:
\beq{mom2} 
0= \sum_{a=1}^3 \sum_{j,k}^{N_a} p^{(a)}_j p^{(a)}_k 
                C^{(a)}_m (\tau^{(a)}_j)^m 
   +\sum_{a=1}^3\sum_j^{N_a}\sum_k^{N_{a+1}} p^{(a)}_j p^{(a+1)}_k 
       \Bigl(\,C^{(a)}_m(\tau^{(a)}_j)^m 
+C^{(a+1)}_m (\tau^{(a+1)}_k)^m \,\Bigr)\ ,
\eeq
where $m$ is an arbitrary integer and the $m$th coefficient 
$C^{(a)}_m$ may depend only on $T_a$. One can add an 
arbitrary number of copies of this identity to $E_G$ with 
different choices of $C^{(a)}_m$. Consider the 
$E_G$ where $G^M_{ab}$ is substituted for $G^{\rm{sym}}$ 
in \eq{gkf2}, and add the identity \eq{mom2} to the $E_G$. 
Then a new set of Green functions can be read from the 
modified $E_G$ as 
\beqa
G'_{aa}(\tau^{(a)}_1,\tau^{(a)}_2)
&=& G^M_{aa}(\tau^{(a)}_1,\tau^{(a)}_2)
+2C^{(a)}_m(\tau^{(a)}_1)^m+\cdots\ , \label{gdash1}\\
G'_{aa+1}(\tau^{(a)}_1,\tau^{(a+1)}_2)
&=&G^M_{aa+1}(\tau^{(a)}_1,\tau^{(a+2)}_2)
+C^{(a)}_m(\tau^{(a)}_1)^m + C^{(a+1)}_m(\tau^{(a+1)}_2)^m
+\cdots\ , \label{gdash2}
\eeqa
where $\cdots$ means the additions of further different copies 
mentioned above. These relations suggest that a variety of 
Green function's representations can be derived starting 
from $G^M_{ab}$. In practice, the three representations listed 
here ($G^{\rm sym}$, $G^M$ $G^J$) are connected in the 
following choices of the $C^{(a)}_m$ coefficients. 
We obtain $G'_{ab}=G^J_{ab}$, if we choose 
\beq{Cj}
C^{(a)}_0=-{1\over2}\Delta T_1T_2T_3\ , \quad
C^{(a)}_1=\Delta {T_1T_2T_3\over T_a}\ ,\quad
\mbox{others}=0\ ,  \label{Cj}
\eeq
and we obtain $G'_{ab}=G^{\rm sym}_{ab}$, if we choose
\beq{*0}
C^{(a)}_1=1\ ,\quad
C^{(a)}_2=-{1\over T_a}(1-\Delta{T_1T_2T_3\over T_a})\ ,\quad
\mbox{others}=0\ .
\eeq
In this way, every possible form is related to the standard 
form by the transformation rules \eq{gdash1} and \eq{gdash2}, 
or in other words, 
by the two-loop momentum constraint formula \eq{mom2}. 

%--------------------------------------------------------------------
\section*{IV. Source constraint}\label{sec4}
\setcounter{section}{4}
\setcounter{equation}{0}
\indent
%--------------------------------------------------------------------

In this section, we discuss what identity in the generating 
functional method should play the same role as the momentum 
conservation constraint \eq{mom2}. 

Let us first recall the computation process from \eq{inter} to 
\eq{inter2}. 
The $w$ integration in \eq{inter} gives rise to the similar 
$\delta$-function divergence as before (cf. Eq.\eq{delta}) in the 
sense of Minkowski formulation, and we then have 
\beqa
Z[J]&=&i\delta\Bigl(\sum_{a=1}^3\int_0^{T_a}\,J^a_\mu(\tau)d\tau
\Bigr)
\prod_{a=1}^3(4\pi T_a)^{-D/2} \nn\\
&\times&\exp[-{1\over2}\sum_{a=1}^3\int_0^{T_a}\int_0^{T_a}
J^a_\mu(\tau) {\tilde G}^{(a)}_{\mu\nu}(\tau,\tau')J^a_\nu(\tau) 
d\tau d\tau'\,]\, I[J] \label{inter3}
\eeqa
with
\beq{IJ}
I[J]=\int dz\prod_{a=1}^3\exp\Bigl[\,
-{1\over4}z^\mu A^a_{\mu\nu}z^\nu 
+z^\nu\int_0^{T_a}J^a_\mu(\tau)R^a_{\mu\nu}(\tau)d\tau\,\Bigr]\ .
\eeq
Here, the $R^a_{\mu\nu}$ given in \eq{AR} is a symmetric tensor, 
however it is not a general property. Rather, the following 
reflection anti-symmetry is general and important: 
\beq{*10}
R^a_{\mu\nu}(\tau)=-R^a_{\mu\nu}(T^a-\tau) \ .
\eeq
Suppose that $J^a_\mu$ behaves as an even or odd function w.r.t. the 
center point $T^a/2$ for the interval $0\leq\tau^{(a)}\leq T_a$; 
i.e.,
\beq{*11}
J^a_\mu(\tau^{(a)})=\pm J^a_\mu(T_a-\tau^{(a)}) \ .
\eeq
Using these properties, we have 
\beq{pro1}
z^\nu\int_0^{T_a}J^a_\mu(\tau) R^a_{\mu\nu}(\tau)d\tau=
\mp z^\nu\int_0^{T_a} R^a_{\nu\mu}(\tau)J^a_\mu(\tau) d\tau\ . 
\label{pro1}
\eeq
By this formula, we perform the Gaussian integral in \eq{IJ}:
\beq{*12}
I[J]=(4\pi)^{D\over2}\mbox{Det}_L^{-{1\over2}}(\sum_a A^a)
\exp[\mp(\sum_a\int_0^{T_a}J^a_\mu\,R^a_{\mu\rho}d\tau)
(\sum_a A^a)_{\rho\sigma}^{-1}
(\sum_a\int_0^{T_a}R^a_{\sigma\nu} J^a_\nu\, d\tau)\,]\ .
\eeq
Again using \eq{pro1}, we can eliminate the complex signature symbol 
\beq{Ires}
I[J]=(4\pi)^{D\over2}\mbox{Det}_L^{-{1\over2}}(\sum_a A^a)
\exp[\,(\sum_a\int_0^{T_a} R^a_{\rho\mu} J^a_\mu d\tau)
(\sum_a A^a)_{\rho\sigma}^{-1}
(\sum_a\int_0^{T_a}R^a_{\sigma\nu} J^a_\nu\, d\tau)\,]
\ .\label{Ires}
\eeq
This result \eq{Ires} holds for any linear combination of even 
and odd $J^a_\mu$ functions. It should be noted that the 
odd source case implies 
\beq{*13}
\int_0^{T_a} J^a_\mu(\tau^{(a)})d\tau^{(a)}=0\ , 
\qquad(a=1,2,3)\ .
\eeq
It means a strong sense 'momentum' conservation 
which is subjected to only one of three internal lines, 
and trivially satisfies 
\beq{Jconst}
\sum_{a=1}^3\int_0^{T_a} J^a_\mu(\tau^{(a)})d\tau^{(a)}=0\ .
\label{Jconst}
\eeq
Mimicking this property, we in general 
impose this identity as the total 'momentum' conservation 
(sum of three lines), as advocated by the $\delta$-function 
in \eq{inter3}.

Now, let us compare the roles of discrete and continuous 
constraints \eq{mom2} and \eq{Jconst} in an example. 
We notice the following term in $G^J_{ab}$ (q.v. \eq{GJ}) 
\beq{kasu}
-{1\over2}(2\tau^{(a)}_1-T_a)(2\tau^{(b)}_2-T_b)
{T_1T_2T_3\over T_a T_b}\Delta
\eeq
and its corresponding term in the generating functional 
\eq{Ires}: 
\beq{kasus}
\ln I[J]= \delta_{\mu\nu}T_1T_2T_3\Delta\sum_a\sum_b\int_0^{T_a}
({\tau^{(a)}\over T_a}-{1\over2})J^a_\mu d\tau^{(a)}
\int_0^{T_b}({\tau^{(b)}\over T_b}-{1\over2})J^b_\nu d\tau^{(b)}
+\cdots\ .
\eeq
If we subtract the $C^{(a)}_m$ terms from \eq{kasu} with the 
choice \eq{Cj}, we obtain $G^M$ as understood from \eq{gdash1} 
and \eq{gdash2}. On the other hand, using the source 
constraint~\eq{Jconst}, we can remove from~\eq{kasus} the linear 
terms in $\tau^{(a)}$ and $\tau^{(b)}$ as well as the constant term:
\beq{*14}
\ln I[J] \approx 
\delta_{\mu\nu}T_1T_2T_3\Delta\sum_a\sum_b\int_0^{T_a}
\int_0^{T_b} {\tau^{(a)}\tau^{(b)}\over T_a T_b} 
J^a_\mu (\tau^{(a)}) J^b_\nu(\tau^{(a)}) d\tau^{(a)} d\tau^{(b)}
+\cdots\ .
\eeq
This manipulation leads to the Green function $G^M_{ab}$ as expected; 
i.e.the removal of the $\tau^{(a)}$ and $\tau^{(b)}$ linear terms 
corresponds to the subtraction of $C^{(a)}_1$ given in \eq{Cj}, 
and the constant term removal to $C^{(a)}_0$. In this way, the 
source constraint \eq{Jconst} plays the same role as the actual 
momentum conservation constraint \eq{mom2} on the kinematical 
factor $E_G$, thus on the wide sense world-line Green functions. 
It is worth noting that the constraint \eq{Jconst} 
is simpler than \eq{mom2}. 

%--------------------------------------------------------------------
\section*{V. Examples in QED}\label{sec5}
\setcounter{section}{5}
\setcounter{equation}{0}
\indent
%--------------------------------------------------------------------

The idea of the source constraint gives a family of equivalent 
world-line Green functions as seen in the previous section. 
This property is useful for identifying different (wide sense) 
Green functions obtained by various computations. In this section, 
we verify its usefulness in more specific cases. The examples 
discussed here is the one-loop photon scatterings in the scalar and 
spinor QED cases. First we discuss the scalar case, and then the 
spinor case. 

The $N$-point function for a complex boson loop is known to be 
given~\cite{str} by the closed path integral of one-dimensional 
bosonic field $x^\mu(\tau)$:
\beq{ampb}
\Gamma_N(p_1,\cdots,p_N)\equiv \int_0^\infty
{dT\over T}\oint{\cal D}x 
(\int_0^T\prod_{j=1}^N d\tau_j d\theta_j{\bar\theta}_j) 
\exp[\, \int_0^T (-{1\over4}{\dot x}^2+J\cdot x)d\tau\,]
\eeq
with the following specific source function
\beq{*15}
J^\mu(\tau)=\sum_{j=1}^N \delta(\tau-\tau_j) 
({\bar\theta}_j\theta_j
\epsilon^\mu_j {\der\over\der\tau_j}+ip^\mu_j)\ ,
\eeq
where $\epsilon^\mu_j$ are photon polarization vectors, and 
$\theta_j$ and ${\bar\theta}_j$ are the Grassmann variables. 
This source is neither an even function nor an odd one in $\tau$, 
and we assume the one-loop  version of the constraint \eq{Jconst} 
to be
\beq{JT}
\int_0^T J^\mu(\tau)d\tau=0\ .
\eeq
This leads to the constraint similar to the momentum conservation 
law
\beq{Jcon}
\sum_{j=1}^N J^\mu_j=0;\qquad
J^\mu_j ={\bar\theta}_j\theta_j\epsilon^\mu_j
{\der\over\der\tau_j} + ip^\mu_j \ .  \label{Jcon}
\eeq
The second term in $J_j$ exactly corresponds to the momentum 
conservation law, while the first term does not vanish in the sum 
at all. In this sense, the present constraint \eq{Jcon} 
assumes a nontrivial conservation law. Let us see how our idea 
works in the following. We perform the path integral~\eq{ampb} 
as the mode integrations with expanding  
\beq{*16}
x^\mu(\tau)=x^\mu_0+\sum_{n>0}x^\mu_n\sin({n\pi\tau\over T})\ ,
\qquad (-\infty\leq x_n\leq\infty) \ .
\eeq
Note that $x_0$ integration diverges as the $\delta$-function 
corresponding to the constraint \eq{JT}. (This is similar to 
the $\delta$-function in Eq.\eq{inter3}). Remember 
that this kind of divergence is usually removed 
by hand (so-called zero mode divergence). 
The resulting expression is then 
\beq{gn}
\Gamma_N(p_1,\cdots,p_N)=\int{dT\over T}({1\over4\pi T})^{D\over2}
(\prod_{j=1}^N d\tau_jd\theta_jd{\bar\theta}_j) 
\exp[\, {1\over2}g_{\mu\nu}\sum_{j,l=1}^N J^\mu_j J^\nu_l
{\tilde G}_B(\tau_j,\tau_l)\,] \ ,
\eeq
where we formally put $g_{\mu\nu}=-\delta_{\mu\nu}$ and
\beqa
{\tilde G}_B(\tau_i,\tau_j)
&=& \sum_{m=1}^\infty{4T\over m^2\pi^2}\sin({\pi m\tau_i\over T})
\sin({\pi m\tau_j\over T}) \\
&=& |\tau_i-\tau_j|-(\tau_i+\tau_j)+2{\tau_i\tau_j\over T}\ .
\eeqa
Here we have used the following formula at the 2nd line of the 
above: 
\beq{formula}
\sum_{m=1}^\infty {\cos(mx)\over m^2}
={1\over4}(|x|-\pi)^2-{\pi^2\over12} \ .
\eeq

Under the constraint \eq{Jcon}, we realize that 
${\tilde G}_B$ in the exponent (the generating kinematical factor) 
in Eq.\eq{gn} behaves as the one-loop Green function \eq{gb} 
exactly, and we thus rederive the same result~\cite{str} 
\beq{*17}
\Gamma_N(p_1,\cdots,p_N)=\int{dT\over T}({1\over4\pi T})^{D\over2}
(\prod_{j=1}^N d\tau_j d\theta_j d{\bar\theta}_j)
\exp[\,-{1\over2}\sum_{j,l=1}^N J_j\cdot J_l G_B(\tau_j,\tau_l)\,]\ .
\eeq
In this example, it is clear that the source constraint helps us 
obtain a correct kinematical factor even if a different (wide sense) 
Green function appears in an intermediate step. 

The similar argument applies to the fermion loop case as well. 
For simplicity, we discuss only the spin part (world-line fermion 
$\psi^\mu(\tau)$), since the bosonic part is essentially the same as 
the above case. The world-line fermion part of the $N$-point amplitude 
is given by~\cite{str}  
\beq{gspin}
{\tilde\Gamma}_N\equiv 
\oint{\cal D}\psi(\prod_{j=1}^N\int_0^T d\tau_j d\theta_j d{\bar\theta}_j) 
\exp[\,\int_0^T(-{1\over2}\psi^\mu\der_\tau\psi_\mu
+\eta^\mu\psi_\mu) d\tau\,]
\eeq
with the source function 
\beq{ytau}
\eta^\mu(\tau) =\sum_{j=1}^N\sqrt{2}(\theta_j\epsilon^\mu_j
+i{\bar\theta}_j p^\mu_j)\delta(\tau-\tau_j)\ .
\eeq
Assuming the source constraint 
\beq{*18}
\int_0^T \eta^\mu(\tau)d\tau=0
\eeq
or equivalently
\beq{Kic}
\sum_{j=1}^N K_j=0\ ,\qquad
K^\mu_j = \sqrt{2}(\theta_j\epsilon^\mu_j+i{\bar\theta}_j p^\mu_j)\ ,
\eeq
and performing the path integral with the mode expansion 
\beq{pmode}
\psi^\mu(\tau)=\sum_{r\in{\bf Z}+{1\over2}} b^\mu_r
\cos({2\pi r\tau\over T})\ ,
\eeq
we obtain (the detail is in Appendix A) 
\beq{gama}
{\bar\Gamma}_N=\prod_{j=1}^N\int d\tau_jd\theta_jd{\bar\theta}_j
\exp[\,{1\over4}\sum_{j,l=1}^NK_j\cdot K_l{\tilde G}_F(\tau_j,\tau_l)\,]
\eeq
with 
\beq{gfw}
{\tilde G}_F(\tau_i,\tau_j)=\mbox{sign}(\tau_j-\tau_i)
+{2\over T}(\tau_i-\tau_j) \ .
\eeq
Under the constraint \eq{Kic}, ${\tilde G}_F$ in the exponent 
plays the same role as the standard fermion Green function
\beq{*19}
G_F(\tau_i,\tau_j)=\mbox{sign}(\tau_j-\tau_i)\ ,
\eeq
and we reproduce the correct answer~\cite{str}
\beq{*20}
{\tilde\Gamma}_N=\prod_{j=1}^N\int d\tau_j d\theta_jd{\bar\theta}_j
\exp[\,{1\over4}\sum_{j,l=1}^NK_j\cdot K_l G_F(\tau_j,\tau_l)\,]\ .
\eeq

%--------------------------------------------------------------------
\section*{VI. Reparametrization type transformation}\label{sec6}
\setcounter{section}{6}
\setcounter{equation}{0}
\indent
%--------------------------------------------------------------------

This section is independent of the previous sections. 
In this section, we consider reparametrizations and transformations 
between the standard two-loop Green functions. As mentioned in 
Sect.II, it is known that $G^{\rm{sym}}$ and $G^{(1)}$ 
are connected by a certain transformation~\cite{HTS}. Here, we 
point out another transformation between them, using 
periodicities of the Green functions, and also discuss 
reparametrizations of $G^{(1)}$ through 
exchanging two of internal lines. 
 
The symmetric Green functions $G^{\rm sym}$ and also 
$G^{(1)}_{33}$ satisfy the following properties of periodicity
\beq{period}
G^{\rm sym}_{ab}(T_a-\tau^{(a)},T_b-\tau^{(b)})
=G^{\rm sym}_{ab}(\tau^{(a)}, \tau^{(b)}), 
\hskip 20pt \mbox{($a$, $b$ = 1, 2, 3)}    \label{period}
\eeq
\beq{*21}
G^{\rm sym}_{aa}(\tau-P_a) =G^{\rm sym}_{aa}(\tau), \quad
G^{(1)}_{33}(\tau-P_{11}) =G^{(1)}_{33}(\tau), 
\eeq
where 
\beq{*22}
P_a = T_a + {T_{a+1}T_{a+2}\over T_{a+1}+T_{a+2}}, \quad
P_{11} = T_3 + G_B(\tau_\alpha,\tau_\beta) ={1\over T\Delta}.
\eeq
Putting $P_3=P_{11}$ with identifying $T_1=T(1-u)$ and $T_2=Tu$, 
we easily find 
\beq{*23}
          u={\vert\tau_\alpha-\tau_\beta\vert\over T},
\eeq
and the necessary relations for the transformation 
between $G^{\rm sym}$ and $G^{(1)}$ \cite{HTS}
\beq{rule1}
T_1=T - \vert \tau_\alpha - \tau_\beta \vert, \quad
T_2= \vert \tau_\alpha - \tau_\beta \vert.        \label{rule1}
\eeq
{}~For later convenience, we assign more concrete notations to 
$\tau_n$ in $G^{(1)}$ on the loop type parameterization:
\beq{xyz}\tau_n=   \left\{\begin{array}{ll} 
x_n & \quad  (\tau^{\ast}<\tau_n)   \\
y_n & \quad  (\tau_n< \tau^{\ast})  \\
z_n & \quad  \mbox{on the internal line $T_3$\ ,} \end{array}\right. 
\label{xyz}
\eeq
where $\tau^{\ast}$ is given below (see Eq.\eq{tstar} and Fig.1). 

%
% Pictex
%
\vspace{8mm}
\begin{minipage}[t]{15cm} 
\begin{center}
%
%        trapic.tex
%
\font\thinlinefont=cmr5
\begingroup\makeatletter\ifx\SetFigFont\undefined
% extract first six characters in \fmtname
\def\x#1#2#3#4#5#6#7\relax{\def\x{#1#2#3#4#5#6}}%
\expandafter\x\fmtname xxxxxx\relax \def\y{splain}%
\ifx\x\y   % LaTeX or SliTeX?
\gdef\SetFigFont#1#2#3{%
  \ifnum #1<17\tiny\else \ifnum #1<20\small\else
  \ifnum #1<24\normalsize\else \ifnum #1<29\large\else
  \ifnum #1<34\Large\else \ifnum #1<41\LARGE\else
     \huge\fi\fi\fi\fi\fi\fi
  \csname #3\endcsname}%
\else
\gdef\SetFigFont#1#2#3{\begingroup
  \count@#1\relax \ifnum 25<\count@\count@25\fi
  \def\x{\endgroup\@setsize\SetFigFont{#2pt}}%
  \expandafter\x
    \csname \romannumeral\the\count@ pt\expandafter\endcsname
    \csname @\romannumeral\the\count@ pt\endcsname
  \csname #3\endcsname}%
\fi
\fi\endgroup
\mbox{\beginpicture
\setcoordinatesystem units <0.66000cm,0.66000cm>
\unitlength=0.66000cm
\linethickness=1pt
\setplotsymbol ({\makebox(0,0)[l]{\tencirc\symbol{'160}}})
\setshadesymbol ({\thinlinefont .})
\setlinear
%
%      Circles
%
\linethickness=14pt
\setplotsymbol ({\thinlinefont .})
\ellipticalarc axes ratio  3.054:3.054  360 degrees 
	from  9.881 19.209 center at  6.826 19.209
\ellipticalarc axes ratio  3.054:3.054  360 degrees 
	from 21.950 19.348 center at 18.895 19.348
%
%     Vertical Lines
%
\linethickness=0.5pt
\setplotsymbol ({\makebox(0,0)[l]{\tencirc\symbol{'175}}})
\putrule from  6.826 22.225 to  6.826 16.192
\putrule from 18.891 22.384 to 18.891 16.351
%
%     Arrows
%
\linethickness=14pt
\setplotsymbol ({\thinlinefont .})
\plot  3.651 19.526  3.821 18.997 /
\plot  3.821 18.997  3.990 19.526 /
\plot  6.657 19.579  6.826 19.050 /
\plot  6.826 19.050  6.996 19.579 /
\plot  9.684 19.526  9.853 18.997 /
\plot  9.853 18.997 10.022 19.526 /
\plot 18.722 19.579 18.891 19.050 /
\plot 18.891 19.050 19.061 19.579 /
\plot 21.738 18.997 21.907 19.526 /
\plot 21.907 19.526 22.077 18.997 /
\plot 15.716 19.526 15.886 18.997 /
\plot 15.886 18.997 16.055 19.526 /

%
%   Tex Objects
%
\put{$\tau^{(1)}$} [lB] at  4.223 19.685
\put{$\tau^{(3)}$} [lB] at  7.620 19.685
\put{$\tau^{(2)}$} [lB] at 10.795 19.685
\put{$T_1$} [lB] at  4.223 18.685
\put{$T_3$} [lB] at  7.620 18.685
\put{$T_2$} [lB] at 10.795 18.685
\put{$x$} [lB] at 16.605 19.685
\put{$\tau^{\ast}$} [lB] at 18.733 22.878
\put{$z$} [lB] at 19.685 19.685
\put{$y$} [lB] at 22.560 19.685
\linethickness=0pt
\putrectangle corners at  2.223 23.381 and 22.860 16.137
\endpicture}

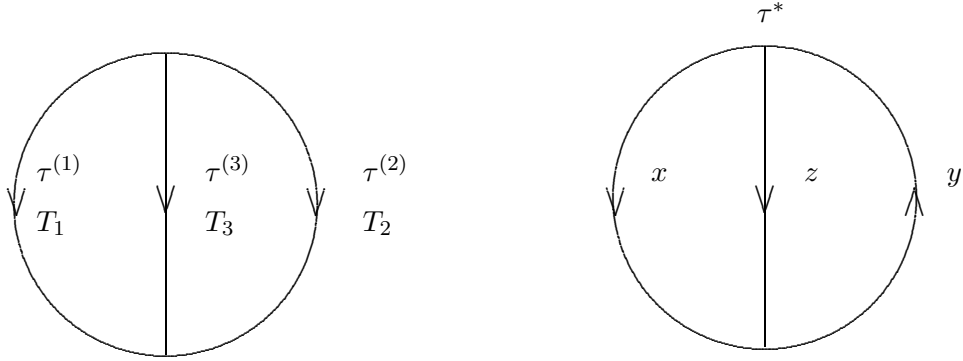
{\bf Figure 1:} The directions of $\tau$ parameters.
\end{center}
\end{minipage}
\vspace{8mm}

\noindent
With these relations, the transformation rule between $G^{\rm sym}$ 
and $G^{(1)}$ is allowed to be expressed as 
\beq{rule2}\left\{\begin{array}{ll} 
\tau_n^{(1)} & =  x_n     -     \tau^{\ast}, \\
\tau_n^{(2)} & =  \tau^{\ast} - y_n,     \\
\tau_n^{(3)} & =  z_n, \end{array}\right.            \label{rule2}
\eeq
where
\beq{tstar}
\tau^{\ast}=
\tau_\alpha\theta(\tau_\alpha-\tau_\beta) 
+ \tau_\beta\theta(\tau_\beta-\tau_\alpha) \ ,
\eeq
\beq{*24}
\tau^{\ast}-T_2 \leq y_n \leq\tau^{\ast}\leq x_n\leq T_1+\tau^{\ast}.
\eeq
We can obtain another transformation 
by combining the property \eq{period} with \eq{rule2}; 
i.e., replacing $\tau^{(a)}\ra T_a-\tau^{(a)}$ 
on the right-hand side in \eq{rule2}
\beq{*25}\left\{\begin{array}{ll} 
\tau_n^{(1)} & =  T_1 - x_n + \tau^{\ast}, \\
\tau_n^{(2)} & =  T_2 + y_n - \tau^{\ast},  \\
\tau_n^{(3)} & =  T_3 - z_n.  \end{array}\right.
\eeq
{}~From these (two-loop) transformations, we can generate an 
infinite number of transformations for the {\it one-loop} case, 
since $\tau^{\ast}$ is reduced to an arbitrary number when the 
both edges of the $z$ line approach each other along the 
fundamental loop (of course, with vanishing $T_3$); for example,
\beq{*26}  \mbox{for} \quad \tau^{\ast}=0: \quad
\left\{ \begin{array}{ll} \tau^{(1)} &= x \\
\tau^{(2)} &= -y \end{array}\right., \quad \mbox{or}\quad
\left\{ \begin{array}{ll} \tau^{(1)} &= T_1 - x \\
\tau^{(2)} &= T_2 + y \end{array}\right.,
\eeq                                              
\beq{onelt}  \mbox{for} \quad \tau^{\ast}=T_2: \quad
\left\{ \begin{array}{ll} \tau^{(1)} &= x - T_2\\
\tau^{(2)} &= T_2 - y \end{array}\right., \quad \mbox{or}\quad
\left\{ \begin{array}{ll} \tau^{(1)} &= T - x \\
\tau^{(2)} &= y \end{array}\right..
\eeq                                              
Although these one-loop transformations are certainly trivial by 
themselves, an interesting deduction is that one can generate a 
set of transformations of $h$-loop Green functions from $(h+1)$-loop 
transformations by setting one of $h$ copies of $\tau^{\ast}$ to 
be an arbitrary value.

As a second application of \eq{rule2}, let us consider some 
reparametrizations of $G^{(1)}_{ab}$; $a,b=0,3$. We show that 
the transformation of the Green functions 
$G^{(1)}_{ab}(\tau_1,\tau_2)$ living on the $zx$-loop (loop made 
of internal lines where the $z$ and $x$ variables are defined) 
into $G^{(1)}_{00}(\tau_1',\tau_2')$ on the $xy$-loop can be 
found through the cyclic permutation symmetry of $G^{\rm sym}$ 
(exchanging $z$-line and $y$-line). Namely, transforming 
$G^{(1)}$ $\ra$ $G^{\rm sym}$ $\ra$ $G^{(1)}$ successively, 
we can read how to transform like
\beqa
G^{(1)}_{00}(x_1,x_2)& &\quad \ra \quad G^{(1)}_{00}({x'}_1,{x'}_2)\\
G^{(1)}_{03}(x,z)& &\quad \ra \quad G^{(1)}_{00}({x'},{y'})\\
G^{(1)}_{33}(z_1,z_2)& &\quad \ra \quad G^{(1)}_{00}({y'}_1,{y'}_2).
\eeqa
Suppose that 
each of $G^{(1)}_{ab}$ on the $xz$-loop is related to 
$G^{\rm sym}_{ij}(\tau^{(i)},\tau^{(j)})$ by the rule \eq{rule2}, 
and that each of $G^{(1)}_{00}$ on the $xy$-loop is related to 
$G^{\rm sym}_{ij}(\tau'^{(i)},\tau'^{(j)})$ by the same rule 
as \eq{rule2}: 
$\tau'^{(1)} =x'-\tau^{\ast}$, $\tau'^{(2)} =\tau^{\ast}-y'$, 
$\tau'^{(3)}=z'$. Putting $\tau'^{(2)}=\tau^{(3)}$ and 
$\tau'^{(3)}=\tau^{(2)}$ (corresponding to the exchange of 
$z$- and $y$-lines), and eliminating $\tau^{(a)}$ and 
$\tau'^{(a)}$ from these transformation rules, we find 
the following transformation rule attributed from the 
exchange between $z$- and $y$-lines:
\beq{repa1}
\left\{
\begin{array}{l}
x'=x            \\
y'=\tau^{\ast}-z \\
z'=\tau^{\ast}-y \\
\end{array}\right. \quad \mbox{and} \quad T_3 \leftrightarrow T_2\ .
\label{repa1}
\eeq
Remember that $\Delta^{-1}$ is invariant in any exchange of $T_a$. 
The simplest check of this rule is the following case:
\beqa
G^{(1)}_{33}(z_1,z_2)
&=&|y'_1-y'_2|-{(y'_1-y'_2)^2\over(T_3+G_B(\tau_\alpha,\tau_\beta))T}
(T_1+T_2)\Bigr|_{T_2\leftrightarrow T_3}\nn\\
&=&|y'_1-y'_2|-{(y'_1-y'_2)^2\over(T_3+G_B(\tau_\alpha,\tau_\beta))T}
(T_3+T(1-u))\nn\\
&=& G^{(1)}_{00}(y'_1,y'_2)\ .
\eeqa
Similarly, we derive another transformation rule from the $yz$-loop to 
the $xy$-loop (exchange of $z$-line and $x$-line):
\beq{repa2}
\left\{
\begin{array}{l}
x'=\tau^{\ast}+z  \\
y'=y \\
z'=x-\tau^{\ast} \\
\end{array}\right. \quad \mbox{and} \quad T_3 \leftrightarrow T_1\ .
\label{repa2}
\eeq

One can organize these two sets of transformations in a unified way: 
Let us express the untransforming (identical) variables 
in \eq{repa1} and \eq{repa2} as
\beq{*27}
\tau = x\theta(\tau-\tau^{\ast})+y\theta(\tau^{\ast}-\tau)\ ,
\eeq
and assign ${\tilde\tau}$ to be the parameter transforming to the 
$z'$ variable
\beq{*28} 
{\tilde\tau}=y\theta(\tau-\tau^{\ast})+x\theta(\tau^{\ast}-\tau)\ ,
\eeq
with considering the transformation
\beqa
G^{(1)}_{00}(\tau_1,\tau_2)& &\quad \ra \quad 
G^{(1)}_{00}(\tau'_1,\tau'_2) 
\label{Gtra1}\\
G^{(1)}_{03}(\tau,z)& &\quad \ra \quad 
G^{(1)}_{00}(\tau',{\tilde\tau}')
\label{Gtra2} \\
G^{(1)}_{33}(z_1,z_2)& &\quad \ra \quad 
             G^{(1)}_{00}({\tilde\tau}'_1,{\tilde\tau}'_2)\ . 
\label{Gtra3}
\eeqa
The above two sets of rules \eq{repa1} and \eq{repa2} are now 
expressed in the compact form 
\beq{*29}
\left\{
\begin{array}{l}
\tau'=\tau  \\
{\tilde\tau}'=\tau^{\ast}-z\mbox{sign}(\tau-\tau^{\ast}) \\
z'=(\tau^{\ast}-{\tilde\tau})\mbox{sign}(\tau-\tau^{\ast})
\end{array}\right.  \quad\mbox{and}\quad T_3 
\quad\leftrightarrow\quad T^{\ast}, \label{rulez}
\eeq
where
\beq{*30}
T^{\ast}=T_2\theta(\tau-\tau^{\ast}) + T_1\theta(\tau^{\ast}-\tau).
\eeq
In order to verify these relations, one should note that $TP_{11}$ 
($=\Delta^{-1}$) is invariant under this transformation rule, 
and also that $T$ transforms as 
\beq{*31} 
T=T^{\ast}+{\tilde T}^{\ast} \quad\ra\quad T_3 +{\tilde T}^{\ast} 
\eeq
where
\beq{*32}
{\tilde T}^{\ast}=T-T^{\ast}=
T_1\theta(\tau-\tau^{\ast}) + T_2\theta(\tau^{\ast}-\tau)\ .
\eeq
It is rather convenient to rewrite the Green function \eq{g01} as
\beq{*33}
G^{(1)}_{03}(\tau,z) = z+\vert \tau-\tau^{\ast}\vert -{1\over TP_{11}}
[z^2T+2z\vert \tau-\tau^{\ast}\vert T^{\ast}
+ (\tau-\tau^{\ast})^2(T_3 +T^{\ast})]          \label{newg01}
\eeq
than considering the original form
\beq{*34}
G^{(1)}_{03}(\tau,z) = G^{(1)}_B(\tau,\tau^{\ast})
+{1\over P_{11}}\left\{ T_3z-z^2-\mbox{sign}(\tau_\alpha-\tau_\beta)
[G_B(\tau,\tau_\alpha) - G_B(\tau,\tau_\beta)] \right\} \ .
\eeq
Applying the transformation \eq{rulez} to 
Eqs.\eq{Gtra1}, \eq{newg01}, and \eq{Gtra3}, we obtain 
\beq{*35}
G^{(1)}_{00}(\tau_1,\tau_2)
=G_B(\tau_1,\tau_2)-{\Delta\over T}T^{\ast}{}^2
(\tau_1-\tau_2)^2 \ ,
\eeq
\beq{*36}
G^{(1)}_{00}(\tau,{\tilde\tau})=G_B(\tau,{\tilde\tau})
-{\Delta\over T}[T(\tau^{\ast}-{\tilde\tau})
             -T^{\ast}(\tau-{\tilde\tau})]^2 \ ,
\eeq
\beq{*37}
G^{(1)}_{00}({\tilde\tau}_1,{\tilde\tau}_2)
=G_B({\tilde\tau}_1,{\tilde\tau}_2)-{\Delta\over T}
{\tilde T}^{\ast}{}^2
({\tilde\tau}_1-{\tilde\tau}_2)^2 \ .
\eeq
These representations are independent of the 
choice of either $\tau^{\ast}=\tau_\alpha$ or $\tau_\beta$, 
and reproduce Eq.(23) of \cite{HTS} for the particular choice 
$\tau^{\ast}=\tau_\beta$ (correcting an error in the literature). 

%\newpage

%--------------------------------------------------------------------
\section*{VII. Conclusions}\label{sec7}
\setcounter{section}{7}
\setcounter{equation}{0}
\indent
%--------------------------------------------------------------------

In this paper, we have investigated two types of the constraints 
on the two-loop kinematical factor and the world-line Green functions. 
One is nothing but the momentum conservation law on external legs, 
and the other is the vanishing constraint on the source term 
integrals along the whole of world-line. Although there is no 
direct connection between two of them, the latter can be regarded 
as a continuous version of the former. Because of the ambiguity 
raised by the constraints, an infinite number of wide sense Green 
functions are in fact possible to take part in the kinematical 
factor exponent. However, we verify that all these Green functions 
can be identified with the standard (restricted) Green functions, 
all of which are reduced from a world-sheet Green function~\cite{RS1}, 
and some of which are related to actual solutions of defining 
differential equations with possessing the rotational invariance 
along the fundamental loop~\cite{SSphi}. 

Conversely, the constraints loosen some imposed restrictions on the 
standard Green functions, and eventually make various evaluations 
and approaches possible. These constraints will be useful for 
analyzing higher loop's world-line Green functions. Especially 
it is clear that the source constraint is much easier to apply 
than the momentum conservation constraint in the multi-loop cases. 
In two-loop Yang-Mills theory, there arises a different Green 
function in the calculation in a constant background 
field~\cite{MSS2}, and the source constraint is actually useful 
to identify the Green function with the standard one in the 
vanishing limit of constant background field (as demonstrated in 
Sections~III and IV). Obviously, the similar thing 
is expected to occur in the multi-loop cases. 
Since expressions of multi-loop Green functions are complicated, 
these constraints will be useful for simplifying the expressions 
or for transforming into convenient forms together with the 
transformation property (suggested below \eq{onelt}). 
It might be interesting to speculate a usefulness of our 
techniques in the thermal world-line cases~\cite{thermo}.

In the final part of the paper, we have considered the 
transformations among the Green functions of standard forms, 
associated with the reparametrizations of the two-loop world-line 
diagram. On the one hand, the form of world-sheet Green function 
is independent of the orderings of two vertices, which join the 
internal line and the fundamental loop. On the other hand, 
the crossing type Green functions \eq{g01} and \eq{gsymaa1}, 
which belong to the type of a correlation between the fundamental 
loop and the internal line, are neither translational invariant 
nor ordering independent. It might be that this gap will be 
filled in some way around by taking account of the discussed 
transformations into the loop type Green function $G^{(1)}_{00}$. 
The crossing type Green functions are necessary in non-Abelian 
gauge theory, and a complexity in the combinatorics problem 
will be caused by this type (similarly to the $\phi^3$ theory 
case~\cite{MSS2}). We hope for a useful parameterization or 
a transformation to overcome these problems.

\appendix

%-------------------------------------------------------------------
\section*{Appendix A. Fermion mode integration}
\setcounter{section}{1}
\setcounter{equation}{0}
\indent
%-------------------------------------------------------------------

We show the derivation of \eq{gama} in this appendix. 
The fermion field \eq{pmode} is an expansion which satisfies 
$\psi(0)=-\psi(T)\not=0$ and $\int_0^T\psi^\mu(\tau)d\tau=0$. 
{}~First, we rewrite
\beq{*38}
H\equiv 
\int_0^T (-{1\over2}\psi^\mu\der_\tau\psi_\mu +\eta^\mu\psi_\mu)d\tau
=I-J\ ,
\eeq
where 
\beq{*39}
I\equiv \int_0^T d\tau_1 d\tau_2 d\tau d\tau'
(\psi(\tau_1)-{1\over2}\eta(\tau)G_I(\tau-\tau_1))
(\delta(\tau_1-\tau_2){-1\over2}{\der\over\der\tau_2})
(\psi(\tau_2)-{1\over2}\eta(\tau')G_I(\tau'-\tau_2))\ ,
\eeq
\beq{JJ}
J\equiv -({1\over2})^3\int_0^Td\tau_1 d\tau_2 d\tau d\tau'
\eta(\tau)\eta(\tau')G_I(\tau-\tau_1)\delta(\tau_1-\tau_2)
{\der\over\der\tau_2}G_I(\tau'-\tau_2)
\eeq
with introducing the function
\beq{GI}
G_I(\tau) ={2\over\pi}\sum_{m\geq1}{1\over m}
\sin({2\pi m\tau\over T})\ , 
\eeq
which satisfies
\beq{*40}
{1\over2} \der_\tau G_I(\tau)= \delta(\tau)-{1\over T}\ ,
\eeq
and 
\beq{*41}
G_I(\tau_1-\tau_2)={\der\over\der\tau_1}G_B(\tau_1,\tau_2)\ .
\eeq
Putting \eq{ytau} and \eq{GI} into \eq{JJ}, 
and performing the integrals, we obtain
\beq{*42}
J=-({1\over2})^3\sum_{j,l}K_j K_l(-{\der\over\der\tau_j})
\sum_{m\geq1}{2T\over\pi^2m^2}\cos({2\pi m(\tau_i-\tau_j)\over T})
\ .
\eeq
Using the summation formula \eq{formula}, we have 
\beq{aj}
J=-({1\over2})^2\sum_{j,l}K_j K_l\Bigl(\, {2\over T}(\tau_j-\tau_l)
-\mbox{sign}(\tau_j-\tau_l)\,\Bigr)\ .
\eeq
Shifting $\psi \ra \psi+{1\over2}\eta G_I$ in the path 
integral~\eq{gspin}, the quantity $I$ is reduced to the free 
integral
\beq{*43}
I\quad\ra\quad -{1\over2}\int_0^T \psi\cdot {\dot \psi}\ ,
\eeq
and this yields nothing but the path integral normalization 
\beq{anorm}
\oint{\cal D}\psi e^{-{1\over2}\int_0^T \psi\cdot{\dot\psi}d\tau}
=1\ .
\eeq
This can be checked by integrating the modes \eq{pmode}. 
Therefore we derive \eq{gama} owing to \eq{aj} and \eq{anorm}
\beq{*44}
{\tilde\Gamma}_N=\oint{\cal D}\psi(\prod_{j=1}^N\int d\tau_j 
d\theta_jd{\bar\theta}_j) e^H
=\prod_{j=1}^N\int d\tau_jd\theta_jd{\bar\theta}_j e^J\ .
\eeq

%
%%%%%%%%%%%%%%%%%%%%%%%%%%%%%%%%%%%%%%%%%%%%%%%%%%%%%%%%%%%%%%%%%%%%%%%%
%                           REFERENCES                                 %
%%%%%%%%%%%%%%%%%%%%%%%%%%%%%%%%%%%%%%%%%%%%%%%%%%%%%%%%%%%%%%%%%%%%%%%%


\begin{thebibliography}{99}
%
%      BGF
%
%\bibitem{BDK} Z. Bern, L. Dixon and D.A. Kosower, hep-ph/9602280, 
%and references are therein.
%%

\bibitem{BK} Z. Bern and D.A. Kosower, \NP{B379}, 451 (1992).
\bibitem{str} M.J. Strassler, \NP{B385}, 145 (1992).

\bibitem{string} A.A. Tseytlin, \PL{B168}, 63 (1986); \\
R.R. Metsaev and A.A. Tseytlin, \NP{B298}, 109 (1988);\\
O.D. Andreev and A.A. Tseytlin, \PL{B207}, 157 (1988).

\bibitem{pao}
P. Di Vecchia, A. Lerda, L. Magnea and R. Marotta, \PL{B351}, 445 (1995);\\
P. Di Vecchia, A. Lerda, L. Magnea, R. Marotta and R. Russo,
 \NP{B469}, 235 (1996).


\bibitem{world} C. Schubert, \ACT{B27}, 3965 (1996);\\
C. Schubert, PRINT-97-273;\\
M.G. Schmidt and C. Schubert, LAPTH-703-98 (hep-th/9810161).

\bibitem{multi} K. Roland, \PL{B289}, 148 (1992); 
SISSA/ISAS 131-93-EP. 

\bibitem{pao3}P. Di Vecchia, A. Lerda, L. Magnea, R. Marotta and R. Russo, 
\PL{B388}, 65 (1996). 
\bibitem{RS1} K. Roland and H.-T. Sato, \NP{B480}, 99 (1996). 
\bibitem{RS2} K. Roland and H.-T. Sato, \NP{B515}, 488 (1998).     
\bibitem{SSqed} M.G. Schmidt and C. Schubert, \PR{D53}, 2150 (1996);\\
        K. Daikouji, M. Shino and Y. Sumino, \PR{D53}, 4598 (1996);\\
        M. Reuter, M.G. Schmidt and C. Schubert, \AP{259}, 313 (1997);\\
        H.T. Sato, \NP{B491}, 477 (1997).
%-------------------------------


\bibitem{SSphi} M.G. Schmidt and C. Schubert, \PL{B331}, 69 (1994). 
\bibitem{HTS} H.-T. Sato, \PL{B371}, 270 (1996).  


\bibitem{MSS} M.G.Schmidt and H.-T. Sato, \NP{B524}, 742 (1998).
\bibitem{MSS2} M.G.Schmidt and H.-T. Sato, to appear in \NP{B} (1999) 
(hep-th/9812229).
\bibitem{thermo} H.T. Sato, to appear in \JMP (1999) (hep-th/9809053);\\
M. Haack and M.G. Schmidt, \EPJ{C7}, 149 (1999);\\
I.A. Shovkovy \PL{B441}, 313 (1998).

%
\end{thebibliography}
\end{document}